# Time desynchronization and Ehrenfest paradox

**A López-Ramos**
Departamento de Física, Universidad de Oviedo
Edificio Este, Campus de Viesques, 33203 Gijón, Spain

E-mail: lramos@uniovi.es

**Abstract.** One of the kinematical effects that give raise to the principle of relativity is time desynchronization of moving clocks. The detailed analysis of this phenomenon is of great importance for leading us to the right (and new) solution of Ehrenfest paradox.

## 1. Introduction

The key point for the resolution of Ehrenfest paradox is the retardation in the transmission of centripetal accelerations. Since this also plays an important role in time desynchronization, we will start with a careful analysis of this last phenomenon.

## 2. Time desynchronization from Lorentz transformation

We shall use, for our reasoning, two clocks, A and B, at rest in an inertial reference frame S' that moves with a uniform velocity with regard to another inertial frame S (Fig. 1).

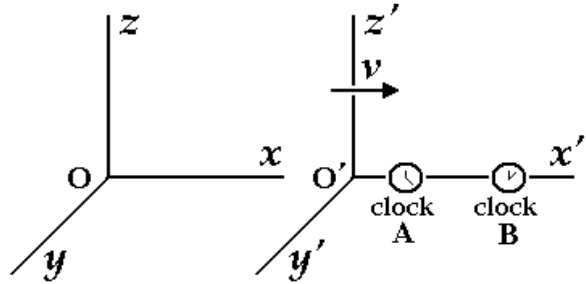

**Figure 1.** Clocks in S'.

The axes and origins for space and time coordinates of both observers have been chosen in the usual way to get the simplest Lorentz transformation and we also assume that clocks A and B have been synchronized in S' reading $t'$:

$$\left.\begin{array}{r}x' = \gamma(x - vt) \\ y' = y \\ z' = z \\ t' = \gamma(t - vx/c^2)\end{array}\right\} \qquad (1)$$

where $v$ is the measured relative velocity and $\gamma = 1/\sqrt{1-(v/c)^2} = 1/\sqrt{1-\beta^2}$

Let two observers in S take photographies of every clock simultaneously in S, being placed next to each clock at that instant. Calling 1 and 2 to the events "taking a photography" of clocks A and B, respectively; we have, from the inverse of the Lorentz transformation formulae involving the times for both observers:

$$t_2 - t_1 = \gamma\left[t'_2 - t'_1 + \frac{v}{c^2}(x'_2 - x'_1)\right] \qquad (2)$$

Due to the fact that events 1 and 2 are simultaneous in S, we have that

$$t_2 - t_1 = 0 \qquad (3)$$

and consequently:

$$t'_2 - t'_1 = -\frac{v}{c^2}(x'_2 - x'_1) \qquad (4)$$

As $t'_1$ and $t'_2$ are, in S', the instants in which the photographies are taken, they coincide with the readings that clocks A and B, respectively, show in the photo films.

The $x'$-coordinates of events 1 and 2 are the same as those of clocks A and B, because they are at rest in S'; so that, finally:

$$t'_1 - t'_2 = \frac{v}{c^2}(x'_B - x'_A) = \frac{v}{c^2}l_0 \qquad (5)$$

being $l_0$ the proper distance between both clocks. Thus, if $x'_B > x'_A$, $t'_1 > t'_2$ and the clock that is behind in the relative movement (clock A) is ahead in time.

It is possible to arrive at the same conclusion starting from two different events 3 and 4: the clocks "show an identical reading" $t_0$ (and, therefore, $t'_3 = t'_4$).

We have now:

$$t_4 - t_3 = \gamma\left[t'_4 - t'_3 + \frac{v}{c^2}(x'_4 - x'_3)\right] = \gamma\frac{v}{c^2}(x'_B - x'_A) \qquad (6)$$

If $x'_B > x'_A$, $t_4 > t_3$ and clock B reads $t_0$ after than clock A. Again the clock that is behind in space is advanced in time.

It must be noticed that $t_4 - t_3$ is not the difference in reading between the clocks, but the elapsed time between the readings of $t_0$ in both clocks from the point of view of S.

During the time interval given by Eq. (6), clock A advances its reading, taking into account time dilation of moving clocks:

$$\frac{1}{\gamma}(t_4 - t_3) = \frac{v}{c^2}(x'_B - x'_A) = \frac{v}{c^2}l_0 \qquad (7)$$

and this will be the difference between the clock readings for S, which is the same result as that in Eq. (5).

## 3. Time desynchronization from synchronizing procedure

Poincaré [1] showed the reason of time desynchronization in 1904: it is due to the fact that a moving inertial observer synchronizes the clocks of his frame by means of light signals *assuming himself at rest*.

Let us analyse from S the synchronization procedure of S'. An observer in S' sends a light signal from clock A to clock B considering himself at rest. He consequently thinks that the velocity of light with regard to him is $c$ and that this signal will need a time interval

$$l_0/c \qquad (8)$$

to arrive at clock B. So, he has, before starting the chronometers, clock A with zero reading and clock B with the reading given by Eq. (8). Then he starts clock A –with zero reading– at the same time that he sends a light signal to start clock B –from the initial reading (8)– by means of a photoelectrical cell in it. In this way, both clocks are synchronized for him.

But from the point of view of an observer in S, the velocity of light is *c* with respect to himself and this observer sees that light has to pursue clock B, that is receding with velocity *v*. Therefore, the time needed by light to arrive at the photoelectrical cell of clock B –for this observer– is, taking also into account length contraction (Fig. 2):

$$\frac{l}{c-v} = \frac{l_0/\gamma}{c-v} \tag{9}$$

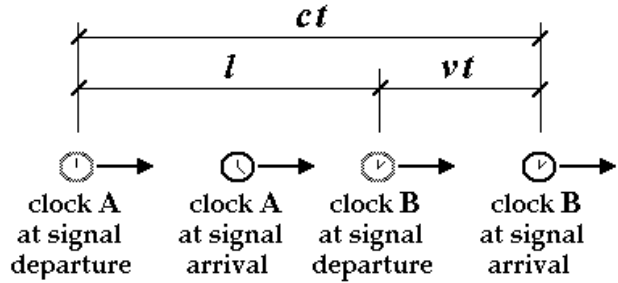

**Figure 2.** Synchronizing procedure.

During the time interval (9) the reading of clock A advances, due to time dilation:

$$\frac{l_0/\gamma^2}{c-v} = \frac{l_0}{c}\frac{1-\beta^2}{1-\beta} = \frac{l_0}{c}(1+\beta) \tag{10}$$

being this its reading when clock B starts. Thus, the reading of clock A is ahead with regard to that of clock B in the difference between Eqs. (10) and (8):

$$\frac{l_0}{c}(1+\beta) - \frac{l_0}{c} = \frac{l_0}{c}\beta = l_0\frac{v}{c^2} \tag{11}$$

which is the same result as that previously deduced from Lorentz transformation.

## 4. Time desynchronization from time dilation

But S' can try to synchronize his clocks in a different way. He starts two chronometers at the same time and at the same place. Then he leaves one of them (clock A) and carries away slowly the other (clock B) to the distance $l_0$.

From the point of view of S, clock B is moving with a greater velocity than that of clock A; suffering, as a consequence of that, a bigger time dilation. Let us call $\Delta v$ the little difference, in the measurements of S, between the velocities of clocks A and B. The time needed to carry clock B to its destination point is, taking into account the length contraction, for S, of the final distance between the clocks in S':

$$T = \frac{l}{\Delta v} = \frac{l_0/\gamma}{\Delta v} \tag{12}$$

During this time interval (12), clock B is moving for S with a velocity $v+\Delta v$, being its rate slowed down in $1/(\gamma+\Delta\gamma)$; while clock A is moving at *v*, being its rate slowed down in $1/\gamma$.

Due to that difference in rate, the reading of clock A is, at the end of the process, bigger than that of clock B in the amount:

$$T\left(\frac{1}{\gamma} - \frac{1}{\gamma+\Delta\gamma}\right) \approx T\frac{\Delta\gamma}{\gamma^2} = \frac{T}{\gamma^2}\Delta(1-\beta^2)^{-1/2} = \frac{T}{\gamma^2}(-1/2)(1-\beta^2)^{-3/2}(-2\beta\Delta\beta) = T\gamma\beta\Delta\beta \tag{13}$$

or, taking into account Eq. (12):

$$\frac{l_0/\gamma}{\Delta v}\gamma\frac{v}{c}\frac{\Delta v}{c}=l_0\frac{v}{c^2} \qquad (14)$$

which, of course, is again the same result as that given by Eq. (5).

It is worth pointing out that this last deduction, completely necessary for the consistence of the principle of relativity cannot be seen before 1923, when it was made by Eddington [2].

In conclusion, in this Section we have seen that when moving at low velocities within a moving inertial frame, clocks read always and everywhere the relativistic local time. As time dilation affects in the same way to persons, local time runs at the same rate as the biological time of all the people moving at non-relativistic velocities within a whole galaxy, with no dependence on the absolute velocity of this galaxy or the individual displacements of each person. Local time gets, in this way, its great physical significance.

## 5. Time desynchronization from mechanical devices

For the consistence of the relativity principle it is also necessary that time desynchronization cannot be avoided by using mechanical devices.

In figure 3, clock B does not work by itself but by means of a rotating rod that joints the main teethed wheels of clocks A and B.

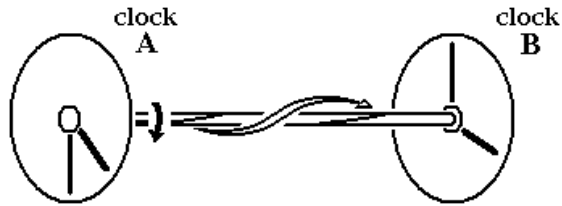

**Figure 3.** Mechanical desynchronization.

If that rod transmits instantly its rotating acceleration, the clocks A and B would be absolutely synchronized and their absolute velocity would be measured by comparing the readings of these clocks and those of clocks synchronized by the methods described in previous Sections.

But the transmission of centripetal acceleration is made by means of electromagnetic and molecular forces between successive atoms in the rod; and, therefore, it cannot be instantaneous. This retardation makes the rod to be twisted, in such a way that its successive transversal sections have a minor angle of rotation as they are farther from the motor teethed wheel of clock A. A straight line parallel to the axis in the rod surface at rest becomes a helix when the rod is put in rotation, as shown in Fig. 3.

We can assume that the transmission of the centripetal acceleration through a rotating rod without translation velocity is made at a velocity $u'=\alpha c$, being $\alpha < 1$ a physical parameter characteristic of the rod, similar to the inverse of the refractive index in a transparent medium.

This will be the case for an observer in S', where the clocks are at rest. So, he thinks that the time needed for the transmission of the rotation between both clocks is

$$\frac{l_0}{\alpha c} \qquad (15)$$

and he advances the initial position of the needle of clock B, with respect to that of clock A, in the amount given by Eq. (15).

But, according to the formulae for the transformation of velocities, the velocity for the transmission of the centripetal acceleration for observer S is:

$$u=\frac{u'+v}{1+u'v/c^2}=\frac{\alpha+\beta}{1+\alpha\beta}c \qquad (16)$$

and the time needed for this signal to reach clock B in S is, taking into account the translation of the rod and its length contraction:

$$\frac{l_0/\gamma}{\left(\frac{\alpha+\beta}{1+\alpha\beta}-\beta\right)c} \tag{17}$$

During this last time interval (17), due to time dilation, the reading of clock A advances:

$$\frac{l_0/\gamma^2}{\left(\frac{\alpha+\beta}{1+\alpha\beta}-\beta\right)c}=\frac{l_0}{c}\frac{1-\beta^2}{\frac{\alpha+\beta-\beta-\alpha\beta^2}{1+\alpha\beta}}=\frac{l_0}{c}\frac{(1-\beta^2)(1+\alpha\beta)}{\alpha(1-\beta^2)}=\frac{l_0}{c}\frac{1+\alpha\beta}{\alpha} \tag{18}$$

which is larger than the initial reading (15) of clock B in

$$\frac{l_0}{c}\left(\frac{1+\alpha\beta}{\alpha}-\frac{1}{\alpha}\right)=\frac{l_0}{c}\beta=\frac{l_0 v}{c^2} \tag{19}$$

leading us once more time to the same result of Eq. (5).

Is it obvious that the same calculation is applied for light signals in transparent media substituting $\alpha$ by the inverse of the refractive index, $1/n$.

All what has been said just to this point confirms that it is not possible to detect the absolute motion by means of electromagnetic, mechanical or combined experiments made within an inertial reference frame. As the principle of relativity says, we can only measure relative velocities.

## 6. Time desynchronization and reciprocity

As it is easy to show, time desynchronization is the key point for the *apparent reciprocity* of the relativistic effects. For instance, Fig. 4 shows the apparent reciprocity of time dilation for a case in which:

$$\Delta t = 30"; \qquad v = c\frac{\sqrt{3}}{2}; \qquad \gamma = \frac{1}{\sqrt{1-(3/4)}} = 2 \tag{20}$$

being time dilation and time desynchronization, respectively:

$$\Delta t_0 = \frac{\Delta t}{\gamma} = \frac{30"}{2} = 15"; \qquad \Delta T = \gamma\frac{v\Delta x}{c^2} = \gamma\frac{v^2}{c^2}\Delta t = 2\frac{3}{4}30" = 45" \tag{21}$$

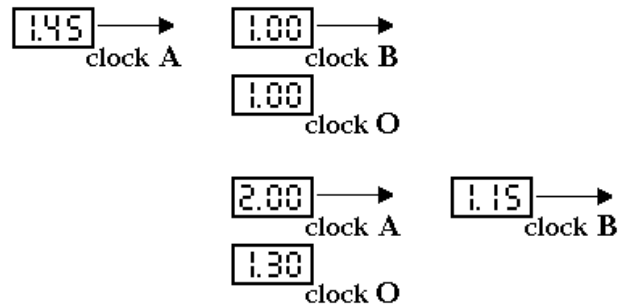

**Figure 4.** Time dilation reciprocity.

In this figure 4, clock O is at rest in an inertial frame S and registers a time interval of 30" between its coincidences with clocks A and B (at rest in a moving frame S'). Despite the fact that these last clocks have increased –meanwhile– their readings in 15" according to S, this observer sees that observers in S' compute 45" more, due to time desynchronization, and obtain that the elapsed time has been one minute.

# 7. Ehrenfest paradox

Ehrenfest or rotating platform paradox arises when a circular platform turns around with high angular velocity. It is supposed that the material of the platform is highly resistant to centrifugal forces. In figure 5, the rim and a diameter of the platform are shown before the rotation, together with another circumference next to the rim, with almost the same perimeter, which will remain motionless during the whole process.

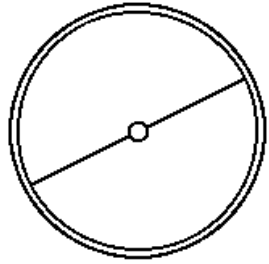

**Figure 5.** Non-rotating platform.

It is generally assumed that the shape of the platform does not change with the rotation, for the inertial observer outside it, and that the platform is not flat for observers rotating with it. The argument given is that measuring rods are contracted when they are parallel to the velocity of the platform points. As a consequence, observers in the platform would obtain a larger value when measuring the rim length, while the measure of the radii is the true value. The geometry of the platform would not be flat for these observers, as the quotient between the rim and the radius would be, then, larger than $2\pi$.

On the other hand, it has been sometimes said that, when the platform is rotating, the length of the rim is contracted according to length contraction formula for the inertial observer outside the platform, because the rim moves along its length with regard to this inertial frame; while the radii are not contracted, as far as they move transversally. The consequence is now that the geometry of the platform would not be flat for the inertial reference frame where it is rotating, as the quotient between the length of the rim and the radius would be, then, smaller than $2\pi$.

There are also authors who say that the perimeter of the rim is contracted only when the rotating material is a ring, but not when it is a whole disk.

All these implausible results come from handling the rotating platform in the same way as inertial frames. When the platform is put into rotation, the transmission, from the axis to the rim, of the forces that provoke the acceleration *cannot be instantaneous*. For this reason, the radii in the platform deform (but keeping themselves in their common plane) at the same time that the length of the rim diminishes, and, contrary to what is usually said, a gap appears between the static and moving circumferences. When the rotation becomes constant, $\omega$, centripetal accelerations *go on being transmitted and keeping the radii bent* while the perimeter of the rotating platform is shortened according to length contraction formula, as is shown in Fig. 6.

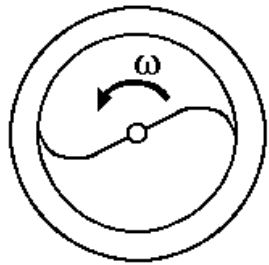

**Figure 6.** Rotating platform.

The real radius, $R$, is the quotient of the perimeter by $2\pi$:

$$R = R_0\sqrt{1 - \omega^2 R^2/c^2} \qquad \Rightarrow \qquad R = R_0\Big/\sqrt{1 + \omega^2 R_0^2/c^2} \qquad (22)$$

and the velocity of the points at the rim ($\omega R$) results smaller than $c$; whatever $\omega$ and $R_0$ are, as it must be.

Of course, for high rotation speed, the deformation due to centrifugal forces, which is much bigger than the relativistic deformation, must be taken into account; but for low rotation speed and large radius, the phenomenon is similar to the deformation due to much lower velocities in very large non-solid massive systems, such as some galaxies (when their masses are not concentrated in the centre and Kepler laws are not applied).

Rods of the same material as the platform change as the platform itself. We can assume that there is a grid of length marks on the platform: they will change under rotation just in the same way as rods upon them. When trying to measure the radius length, rods will be placed by observers in the platform along the bent radius in Fig. 6 (these rods can be fixed along the radius from the beginning). If these observers do not correct their measurements, they would measure the initial values for lengths of both perimeter and radius, in such a way that they also obtain $2\pi$ as the quotient between these two magnitudes. On the other hand, if they correct the results of their measurements, conscious of their own rotation and their non-inertial status, these observers would get the same values as observers outside the platform. In all cases and for every observer, the rotating platform remains flat.

If the transmission of the acceleration is made from the rim to the centre, the radii bend in the opposite direction. In such a case, the radii represented in Fig. 6 would correspond to a clockwise rotation.

Although it is not possible to get platforms rotating at relativistic velocities, the issue is of great theoretical interest, as can be seen from the controversy about it [3-14].

## 8. Conclusions

When analysing the Ehrenfest paradox is necessary taking into account that a rotating platform is not an inertial reference frame, but a frame in which accelerations are being transmitted, in such a way that a deformation is always present in it, due to the non-instantaneous character of the acceleration transmission.

On the other hand, the previous analysis of time desynchronization shows the necessity of this deformation in a rotating system. Without it, the principle of relativity would be refuted.

Finally, the analysis of the deformation shows that space remains flat for all observers: inertial or not inertial.